\documentclass[floats,prb,twocolumn]{revtex4}

\usepackage{graphicx}
\usepackage{amsfonts}
\usepackage{amssymb}
\usepackage{amsmath}

\def\der#1#2{{\partial#1 \over \partial#2}}
\def\be{\begin{equation}}
\def\ee{\end{equation}}
\def\ba#1{\begin{array}{#1}}
\def\ea{\end{array}}
\def\bn{\begin{enumerate}}
\def\en{\end{enumerate}}

\def\rr{\right}
\def\l{\left}

\def\H{\mathcal{H}}
\def\summ{\sum\limits}
\def\intt{\int\limits}

\def\G{\Gamma}

\def\Qg{Q_{\Gamma}}

\def\ag{a_{\Gamma}}
\def\bg{b_{\Gamma}}

\begin{document}

\title{Infinite Randomness Phases and Entanglement Entropy of the Disordered Golden Chain}
\author{L. Fidkowski}
\author{G. Refael}
\affiliation{Department of Physics, Institute for Quantum Information, California Institute of
  Technology, MC 114-36, Pasadena, CA 91125}
\author{N.~E.~Bonesteel}
\affiliation{Department of Physics and National High Magnetic Field Laboratory,
Florida State University, Tallahassee, FL 32310}
\author{J.~E.~Moore}
\affiliation{Department of Physics, University of California,
Berkeley, CA 94720}
\affiliation{Materials Sciences Division,
Lawrence Berkeley National Laboratory, Berkeley, CA 94720}
\begin{abstract}
Topological insulators supporting non-abelian anyonic excitations
are at the center of attention as candidates for topological
quantum computation. In this paper, we analyze the ground-state
properties of disordered non-abelian anyonic chains. The
resemblance of fusion rules of non-abelian anyons and real space
decimation strongly suggests that disordered chains of such anyons
generically exhibit infinite-randomness phases. Concentrating on
the disordered golden chain model with nearest-neighbor coupling,
we show that Fibonacci anyons with the fusion rule
$\tau\otimes\tau={\bf 1}\oplus \tau$ exhibit two infinite-randomness phases: a
random-singlet phase when all bonds prefer the trivial fusion channel,
and a mixed phase which occurs whenever a finite density of bonds
prefers the $\tau$ fusion channel. Real space RG analysis shows that
the random-singlet fixed point is unstable to the mixed fixed
point. By analyzing the entanglement entropy of the mixed phase, we find its
effective central charge, and find that it {\it increases} along the RG
flow from the random singlet point, thus ruling out a c-theorem for the
effective central charge.

\end{abstract}

\maketitle

\section{Introduction}

\begin{figure}
\includegraphics[width=6cm]{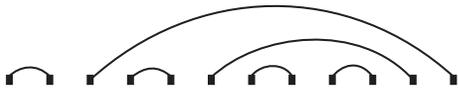}
\caption{Sample random singlet ground state of a spin-1/2
  antiferromagentic chain.  Each line represents a singlet.  Singlets form in a random fashion, mostly between nearest neighbors, but occasionally between distant sites giving rise to an average correlation that decays algebraically. \label{ff8}}
\end{figure}

One of the most interesting frontiers of physics is the behavior
of interacting, many body, quantum systems. Such systems are
particularly challenging and rich when considered in low
dimensionalities, and in the presence of disorder. A common
platform for the discussion of collective behavior is a quantum
magnet. Already in one dimension, where the behavior of quantum
magnets is supposed to be the simplest, surprises emerged: the
Haldane gap of integer spin chains \cite{Haldane,AKLT}, and, more
importantly for this work, the random singlet phase
\cite{DSF94,DSF95}. The latter describes the ground state of a
disordered spin-1/2 Heisenberg chain, where the spins pair up in
and form singlets in a random fashion (fig. \ref{ff8}). Most of
the singlets connect near neighbors, but some are very long
ranged, and lead to algebraically decaying average
correlations. The random singlet phase is the first known example
of the  infinite-randomness paradigm of 1d random systems.
Contrary to the quantum scaling in pure systems, where $1/E\sim
L^{z}$, infinite randomness systems obey the scaling $|\ln E| \sim
L^{\psi}$, and exhibit many other intriguing properties.

Another paradigm for interacting quantum matter is the fractional
quantum Hall system. In addition to robust fractionally-charged
excitations, Hall bars with electronic densities tuned to special
fractions, such as $\nu=5/2$, and $\nu=12/5$ are expected to exhibit
non-abelian quasi-particles excitations and defects \cite{ReadGreen,
  MooreRead}, which may be used to realize a topologically protected
qubit \cite{DSFN, FKLW}, but more importantly, provide an example of a completely
new type of quantum matter. Non-abelian anyons display a
remarkable feature: the dimension of the Hilbert space spanned by
$N$ non-abelions grows asymptotically as $D^N$, where $D$, the
quantum dimension of the non-abelion, is irrational. This is a
consequence of the so-called fusion rules of the non-abelions. In
this paper we will investigate the properties of one-dimensional
disordered systems composed of non-abelions, concentrating on the
case of the Fibonacci anyons, for which the allowed  values of the
so-called topological charge can be either 1 or $\tau$.

The investigation of the so-called Fibonacci chain has so far
concentrated on the analysis and phases in the translationally
invariant case \cite{Qpeople1, Qpeople2}.  It turns out that the
system is exactly solvable by mapping to an RSOS model, and is
described at low energies by a minimal model conformal field
theory with central charge $c=\frac{7}{10}$ in the
antiferromagnetic case (favoring fusion into the trivial channel)
and $c=\frac{4}{5}$ in the ferromagnetic case (favoring fusion
into the $\tau$ channel).   The richness of this example stems
from the unique structure of the Hilbert space of a system
comprising non-abelian anyons.  As we argue below, however, an
important insight is that {\it the construction of the Hilbert
state of a random non-abelian chain is analogous to the
construction of the ground state and low-lying excitation spectrum
of a spin-chain}. Furthermore, contrary to spin chains comprising
garden-variety spins, the Hilbert space structure of non-abelian
chains guarantees the appearance of an infinite randomness scaling
in the presence of disorder. The exotic nature of the non-abelions
suggests that the infinite-randomness phases they will exhibit
will be new, and perhaps even expand our dictionary of
infinite-randomness universality classes, currently limited to the
permutation symmetric sequence \cite{DamleHuse, MonthusReview, hoyos}.

Another interesting aspect of infinite-randomness phases is their
entanglement entropy. The bipartite entanglement entropy of a pure
spin chain at criticality scales logarithmically with its size,
and is proportional to the central charge of the conformal field
theory describing the critical point \cite{Holzhey94,Calabrese04,
Vidal03}.  Furthermore, the central charge, and therefore also the
entanglement entropy of a pure spin chain, obeys the Zamolodchikov
c-theorem: it must decrease along renormalization group flow
lines. Random spin chains also have an entropy that scales
logarithmically with size, and with a universal coefficient that
we identify as an effective central charge
\cite{RefaelMoore2004,RefaelMoore2007, Laflorencie,santachiara}.
An outstanding question has been whether the c-theorem applies to
renormalization group flows between infinite randomness fixed
points of the random chains. The evidence so far has been limited,
since the overwhelming majority of entanglement entropy
calculations were done in the random singlet phase of various
systems. The only exception so far has been the entanglement
entropy at the critical point between the Haldane phase and the
random singlet phase of a spin-1 random antiferromagnet, where the
effective central charge indeed decreases along real-space RG flow
lines \cite{RefaelMoore2007}.  In this paper, we find a real-space
RG flow along which the effective central charge increases, thus
violating any conjectured $c$-theorem for flows between strong
randomness fixed points.

The first analysis of the random Fibonacci chain \cite{YB}
concentrated on the random antiferromagnetic Fibonacci chain.  A
random singlet phase was found with an effective central charge
reflecting the quantum dimension of the Fibonacci anyons:
$c_{eff}=\ln \tau$ with $\tau=\frac{1+\sqrt{5}}{2}$ being the
golden ratio.  In this paper we extend the analysis to the
completely disordered Fibonacci chain, which contains both AFM
(antiferromagnetic, favoring fusion into a singlet) as well as FM
(ferromagnetic, favoring fusion into a $\tau$) bonds - we will
refer to it as the {\it mixed} Fibonacci chain.  We find that the
random singlet phase is unstable to FM perturbations, and flows to
a stable fixed point which, at low energy, is described by an
equal mixture of FM and AFM bonds, with identical infinite
randomness universal strength distributions. We calculate the
entanglement entropy of this new fixed point, and show that it is
larger than that in the unstable AFM random-singlet point.  We
thus have an example of a flow between two infinite randomness
fixed points along which the effective central charge increases.

In the rest of the paper we will describe our analysis of the
random Fibonacci chain. In Sec. \ref{bkg} we review the real-space
renormalization group and the Hilbert space and Hamiltonian of the
Fibonacci chain.  In Sec. \ref{irf} we define the model, and
discuss the decimation rules necessary for a real-space RG
analysis. The decimation steps are then used for the calculation
of flow equations for the disorder distribution. We use these to
derive the phase diagram and investigate the stability of the
fixed points found from the flow equations. Sec. \ref{ee} will
describe the entanglement entropy calculation for the random
Fibonacci chain. Before concluding, we will discuss the
correspondence between the construction of the ground state of a
random spin chain, and the construction of the Hilbert space of a
chain of non-abelian anyons. This provides the basis for further
investigation of other kinds of non-abelian chains, such as the
full $SU(2)_k$ sequence.

\section{Background \label{bkg}}

\subsection{Real Space Renormalization Group}

To find the ground state of disordered spin chains, Ma and
Dasgupta introduced the strong disorder real-space renormalization
group \cite{MaDas1979,MaDas1980}.  The random spin $1/2$
Heisenberg model provides the simplest example for this method.
The model is given by: \be H=\sum_i J_{i, i+1} \, {\bf S}_i \cdot
{\bf S}_{i+1}, \ee where the couplings $J_{i,i+1}>0$ are positive
and randomly distributed.  Note that, as far as the Hilbert space
is concerned, we have for two neighboring sites \be \label{tp}
\frac{1}{2} \otimes \frac{1}{2} = 0 \oplus 1, \ee and the
interactions in the Hamiltonian simply give an energy splitting
between the two representations on the right hand side.  The
procedure now is to pick the largest $J_{i,i+1}$, which
effectively truncates the excited triplet and leaves the ground
state in a singlet, and do perturbation theory around that state.
Quantum fluctuations then induce an effective coupling according
to the so-called Ma-Dasgupta rule \cite{MaDas1979, MaDas1980}: \be
J_{i-1,i+2} = \frac{J_{i-1,i} J_{i+1,i+2}}{2 J_{i,i+1}} \label{md}
\ee So sites $i$ and $i+1$ are decimated and replaced with an
effective interaction between $i-1$ and $i+2$.  Iteration of this
procedure produces bonds on all length scales.  This is the random
singlet ground state.

A quantitative description is obtained by tracking the RG flow of the
coupling distribution.  It is useful to employ logarithmic couplings
\cite{DSF94}:
\be
\beta_{i,i+1} = \ln \frac{\Omega}{J_{i,i+1}}
\ee
where $\Omega = \text{max}_i \, J_{i,i+1}$.  In these variables the
Ma-Dasgupta rule (\ref{md}) reads
\be
\beta_{i-1,i+2} = \beta_{i-1,i} +
\beta_{i+1,i+2}
\ee
(up to an additive constant of $\ln 2$ which can be safely neglected).  As the couplings get decimated $\Omega$
decreases.  It is convenient to define the RG flow parameter as
\be
\Gamma = \ln \frac{\Omega_0}{\Omega}
\ee
where $\Omega_0$ is the maximal coupling of the bare Hamiltonian.  Let
$P_\Gamma(\beta)$ be the distribution of couplings.  We can derive a
flow equation for $P_\Gamma(\beta)$ by decimating the couplings in the
infinitesimal interval $\beta = [\,0, d \Gamma \,]$ and seeing how
their probabilistic weight is redistributed.  We obtain
\begin{eqnarray}
& &\frac{d}{d \Gamma} P_\Gamma (\beta) = \der{P_\Gamma}{\beta} +  \\
& & P(0) \int_0^\infty d \beta_1 \int_0^\infty d \beta_2
  \delta(\beta-\beta_1-\beta)  P_\Gamma(\beta_1)  P_\Gamma(\beta_2)
  \nonumber
\end{eqnarray}
The first term comes from the overall
change of scale, and the second from the Ma-Dasgupta rule.  These
equations have a solution
\be
P_\Gamma (\beta) = \frac{1}{\Gamma} e^{-\beta / \Gamma}
\ee
which is an attractive fixed point to essentially all physical initial
configurations.  This solution permits us to read off features of the
random singlet phase; for example one can with a little more work
derive the energy-length scaling relation:
\be
L^{1/2} \sim \Gamma = \ln \, (1/E).
\ee
which thus has the exponent:
\be
\psi=1/2.
\ee

\subsection{Hilbert Space and Hamiltonian of the Fibonacci Chain}

We now construct the Hilbert space and Hamiltonian of the
Fibonacci chain.  The system is modeled as a chain of non-abelian
anyons carrying the non-trivial topological charge $\tau$.
Heuristically, we want the property that \be \label{ttp} \tau
\otimes \tau = 1 \oplus \tau \ee which states that the Hilbert
space of two neighboring $\tau$'s is the direct sum of a trivial
component and another copy of $\tau$.  This unusual property
immediately prevents us from describing the Hilbert space as a
tensor product of local degrees of freedom.  Indeed, a naive
interpretation of the tensor product in (\ref{ttp}) would give the
dimension of the space $\tau$ to be the golden ratio, an
irrational number.  This problem is resolved by the adoption of
the machinery of {\it truncated} tensor products of
representations of $SU(2)$ at level $k$, but rather than
developing it here we instead give two elementary constructions of
the Hilbert space.  We note, however, the analogy between
(\ref{tp}) and (\ref{ttp}); indeed the Hamiltonian, defined below,
will simply yield an energy splitting between the two
representations on the right hand side of (\ref{ttp}).

\begin{figure}
\includegraphics[width=7cm]{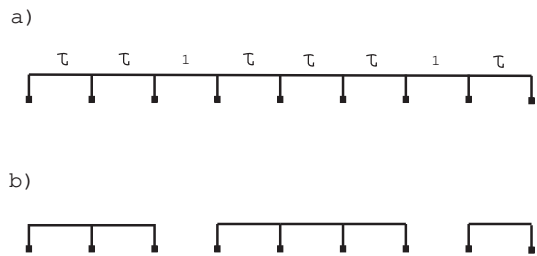}
\caption{a) A state in the Hilbert space: the labels $1$ and $\tau$ specify the total topological charge of all the sites to the left (or equivalently right) of the bond, with the fusion rules obeyed at each trivalent node b) The same state in graph notation - we only draw the $\tau$'s. \label{ff2}}
\end{figure}

The simplest way to construct the Hilbert space is to define basis
states by labeling each link between two $\tau$'s with a $1$ or
$\tau$, with the constraint that one is not allowed to have two
consecutive $1$'s - fig. \ref{ff2}(a).  The dimension $D_N$ of the
Hilbert space for $N$ sites then follows the Fibonacci recursion
\be
D_N = D_{N-1} + D_{N-2}
\ee
which is solved by $D_N \sim \tau^N$.  Thus there are $\tau \simeq
1.618$ ``degrees of freedom" on each site (note: we use $\tau$ to
denote the nontrivial topological charge and the value of the golden
mean, as well as the corresponding representation of $SU(2)_k$ where
appropriate).

\begin{figure}
\includegraphics[width=6cm]{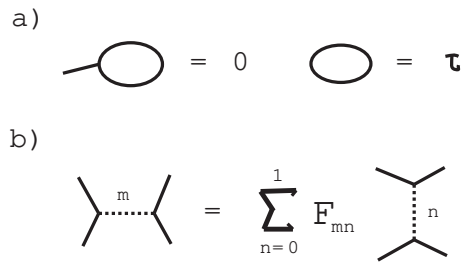}
\caption{The F-matrix relations.  a) graphs that can be disconnected by cutting one edge are equal to $0$ (the no tadpole condition) and disconnected loops are worth $\tau$.  b)  local reconnection rules are given by the $F$-matrix, defined in the text.  Here $m$ and $n$ are binary variables equal to either $\tau$ or $1$ - i.e., the link is either there or not \label{ff3}}
\end{figure}

\begin{figure}
\includegraphics[width=8.5cm]{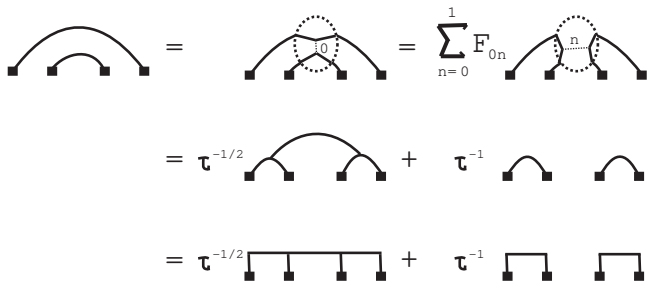}
\caption{The trivalent graph on the left hand side expressed in terms of the link basis. \label{ff10}}
\end{figure}

While this link description of the Hilbert space is most convenient
computationally, there is an equivalent but more abstract one that is
useful in defining the Hamiltonian and carrying out the real space RG
procedure.  In this abstract description the Hilbert space is defined
as the set of all trivalent graphs with endpoints at the $N$ nodes,
modulo the $F$-matrix relations - see fig.\ref{ff3} - where the $F$
matrix is
\begin{equation}
F = \begin{bmatrix}
\tau^{-1} & \tau^{-1/2} \\
\tau^{-1/2} & -\tau^{-1}
\end{bmatrix}
\end{equation}
The edges of the graph represent nontrivial topological charge $\tau$
and trivalent vertices represent the fusion of two $\tau$'s into
another $\tau$.  To relate this graphical picture to the link basis,
note that the link basis states can be viewed as trivalent graphs, as
in fig. \ref{ff2}(b), and any other trivalent graph can be reduced to
a superposition of these using $F$-matrix moves (for example, see
fig. \ref{ff10}).  The inner product of two graphs is defined by
reflecting one of the graphs and concatenating it with the other along
the $N$ nodes.

\begin{figure}
\includegraphics[width=6cm]{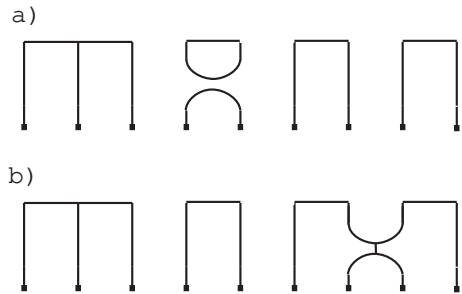}
\caption{a) Projection operator $P_i^A$ on a pair of sites.  b) Projection operator $P_i^F$ on a different pair of sites. \label{ff4}}
\end{figure}

We now define the Hamiltonian.  There are two kinds of nearest
neighbor interactions we consider: we can either project onto total
topological charge $\tau$ of the pair, in which case we refer to the
interaction as ferromagnetic (F), or onto the trivial charge $1$ -
this interaction is antiferromagnetic (A).  The F and A designations
are by analogy with the spin $1/2$ case, where antiferromagnetic
interactions favor a singlet, which has trivial spin, and
ferromagnetic interactions favor non-zero total spin
\be
\H=\summ J_i(1-P_{i}^{\Sigma_i}),
\label{hfib}
\ee
where for each site $J_i$ is a positive random number with a given
distribution, and $\Sigma_i=A$ for the Hamiltonian describing the AFM
fixed point, while $\Sigma_i=F,\,A$ at random for the Hamiltonian
describing the mixed FM/AFM fixed point.  Here $P_i^{F/A}$ are the
projectors onto the ferromagnetic and antiferromagnetic sectors of the
pairs of sites.  These projection operators are defined graphically in
fig. \ref{ff4}.  They can also be viewed in the link basis if we apply
$F$-matrix rules to decompose the result of the concatenations in
fig. \ref{ff4}.

\begin{figure}
\includegraphics[width=6cm]{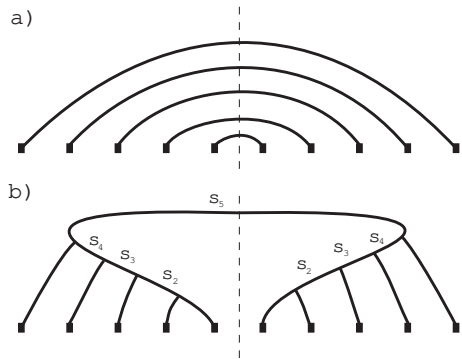}
\caption{A rainbow diagram.  The partition bond is in the middle. \label{ff5}}
\end{figure}

As an illustration of the complicated nature of the Hilbert space, we
analyze the ``rainbow" state that arises in the entanglement entropy
calculation for the AFM chain \cite{YB} - see fig. \ref{ff5} (a).  If
we assume the partition bond (i.e., the bond that divides the system
into two subsystems) to lie in the middle of the chain, then this
state turns out to be quite entangled.  We will be more precise later,
but the idea is that to compute the entanglement entropy, we want to
use $F$-matrix moves to write the state as a superposition of states
shown in fig. \ref{ff5} (b).  This way we push all the nontrivial
parts of the graph into one of the two halves, and the entanglement
can be read off from the coefficients of the new states.  Carrying
this out, Ref. \onlinecite{YB} showed that for large number $N$ of
singlets, the entropy is asymptotically $N \log_2 \tau$.  Thus the
asymptotic contribution of each singlet is $\log_2 \tau$.  Note,
however, that when $N$ is small, there are deviations from this form.
In particular, when $N=1$, so that we have only one $\tau$ anyon in
each half, the Hilbert space has only one state and so the state
counting entropy is $0$.  These sorts of subtleties will be treated
carefully when we do our entropy calculation for the mixed FM/AFM
fixed point.

\section{Infinite randomness fixed points of the golden chains \label{irf}}

\subsection{Ma-Dasgupta decimation rules}

The graphical description of the Hilbert space in terms of
trivalent graphs shows how the real space RG method can be
generalized to the case of the mixed Fibonacci chain, containing
both FM and AFM interactions.  As before one first picks the
largest coupling $J_i$ in eqn. (\ref{hfib}) and assumes that it
localizes a state on $i$ and $i+1$ with total topological charge
either $1$ or $\tau$, depending on whether the interaction is AFM
or FM.  Graphically this localization is just a restriction to
graphs that have a singlet spanning the two sites (AFM case) or
graphs that have the two $\tau$'s at $i$ and $i+1$ fuse into
another $\tau$ (FM case).   Again the state of the two sites is
perturbed by the other two bonds connecting these sites to the
rest of the chain.

To study the effect of this perturbation consider the Hamiltonian
(\ref{hfib}) acting on four tau particles with site labels 1
through 4. Using the fact that $P^F_i = 1 - P^A_i$ this
Hamiltonian can, up to an irrelevant constant, be taken to be
\begin{eqnarray}
H = -J_1 P^A_1 - J_2 P^A_2 - J_3 P^A_3,\label{4spin}
\end{eqnarray}
where now the sign of a given $J_i$, connecting particles at sites
$i$ and $i+1$, can be positive or negative, corresponding to AFM
or FM bonds, respectively. We then assume that $J_2$ is the
highest energy bond, with $|J_2| \gg |J_1|,|J_2|$, and write
(\ref{4spin}) as $H = H_0 + H^\prime$ where $H_0 = -J_2 P^A_2$ is
the ``unperturbed" Hamiltonian and $H^\prime = -J_1 P^A_1 - J_3
P^A_3$ is the perturbation.

First consider the case of decimating an AFM bond for which $J_2 >
0$. The two degenerate ground states of $H_0$ will have a singlet
connecting particles 2 and 3 (i.e., these two particles will have
total topological charge 1) while particles 1 and 4 can combine to
either have topological charge 1 or $\tau$.  We denote these two unperturbed states
$|\psi_1\rangle$ and $|\psi_\tau\rangle$.  Since the total
topological charge of the four particles is a ``good" quantum
number, the perturbation $H^\prime$ will lead to an energy
splitting, $J_{eff}$, between the state of these four particles
with total topological charge 1 and the state with total
topological charge $\tau$. This energy splitting can then be
described by a new Hamiltonian $H_{eff} = - J_{eff} P^A_1$, where
now $P^A_1$ acts on particles 1 and 4.  It is straightforward to
compute $J_{eff}$ using second order perturbation theory with the
result,
\begin{eqnarray}
J_{eff} = \frac{|\langle \psi_1| H^\prime P^F_2 H^\prime
|\psi_1\rangle|}{J_2} - \frac{|\langle \psi_\tau| H^\prime P^F_2
H^\prime |\psi_\tau\rangle|}{J_2}.\label{pert}
\end{eqnarray}
In this expression the FM projection operator $P^F$ projects
$H^\prime |\psi_{1,\tau}\rangle$ onto the excited Hilbert space of
the unperturbed Hamiltonian $H_0$ with energy $J_2$ above the
ground state.

Using the techniques described in the previous section the matrix
elements appearing in (\ref{pert}) can be evaluated to find
\begin{eqnarray}
|\langle \psi_1| H^\prime P^F_2 H^\prime |\psi_1\rangle|^2 &=&
(J_1 + J_3)^2 \frac{1}{\tau^2} \nonumber \\ &-& \left((J_1 + J_3)
\frac{1}{\tau^2}\right)^2,
\\
|\langle \psi_\tau| H^\prime P^F_2 H^\prime |\psi_\tau\rangle|^2
&=& (J_1^2 + J_3^2) \frac{1}{\tau^2} \nonumber \\ &-& \left((J_1 +
J_3) \frac{1}{\tau^2}\right)^2.
\end{eqnarray}
It then follows that
\begin{eqnarray} \label{MD}
J_{eff} = \frac{2}{\tau^2} \frac{J_1 J_3}{J_2}.
\end{eqnarray}
Thus we see that when an AFM bond is decimated the usual
Ma-Dasgupta rule holds. The value of the coefficient $2/\tau^2$ is
not significant except for the fact that, because it is less than
1, $J_{eff}$ will always be less than $J_2$.  As for the usual
Ma-Dasgupta rule, the resulting interaction will be AFM if $J_1$
and $J_3$ have the same sign, and FM if $J_1$ and $J_3$ have
opposite signs.

Next consider the case of a FM bond for which $J_2<0$ in
(\ref{4spin}). In this case the two tau particles connected by
$J_2$ fuse to form a cluster with topological charge $\tau$. The
tau particles on either side of this cluster will then interact
with it, but with modified interaction strengths $\tilde J_1$ and
$\tilde J_3$.

\begin{figure}
\includegraphics[width=5cm]{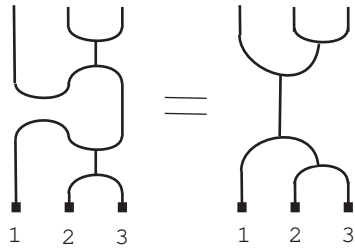}
\caption{Particles $2$ and $3$ are decimated ferromagnetically.
The effect on an AFM projection acting on $1$ and $2$ is simply to
turn it into an FM projection.  For an AFM projector (FM
projector) there is a numerical factor associated to the graphical
representation above of $\tau^{-1}$ ($\tau^{-1/2}$).  Combining
these, we get $\tau^{-1} (\tau^{-1/2})^2 = \tau^{-2}$, of which
$\tau^{-1/2}$ gets absorbed in the normalization for the new FM
projector and another $\tau^{-1/2}$ goes into the normalization of
the wavefunction, leaving $\tau^{-1}$.  Thus $P^A$ on sites $1$
and $2$ turns into $\frac{1}{\tau} P^F$.  A similar graphical
argument shows that $P^F$ turns into $\frac{1}{\tau} P^A$.
\label{ff14}}
\end{figure}

To compute these modified interactions, consider particle $1$,
which is coupled to the newly formed cluster through the ``bare"
interaction $J_1$.  To see the effect of the decimation on any
operator $O$ on the pair of particles $1$ and $2$, we simply
project this operator down to the decimated subspace:
$O_{\text{new}} = P^F \, O \, P^F$ where $P^F$ acts on particles
$2$ and $3$.  Composing the operators graphically in fig.
\ref{ff14} we see that $P^F$ turns into $P^A$ and vice versa, with
an extra factor of $1 / \tau$.

Another way to see this is as follows.  There are two possible
states for the tau particle at site 1 and the $\tau$ cluster
formed by particles 2 and 3 --- the total topological charge of
all three particles can be either 1 or $\tau$. Again we denote
these two states $|\psi_1\rangle$ and $|\psi_\tau\rangle$. In this
case the effective interaction can be computed using {\it first}
order perturbation theory in $H^\prime = -J_1 P^A_1$ with the
result
\begin{eqnarray}
\tilde J_1 = \langle \psi_1 | J_1 P^A_1 |\psi_1 \rangle - \langle
\psi_\tau | J_1 P^A_1 |\psi_\tau\rangle.
\end{eqnarray}
The calculation of these matrix elements is again straightforward and we find that
\begin{eqnarray}
\langle \psi_1 |P^A_1| \psi_1\rangle &=& 0,\\
\langle \psi_\tau |P^A_1|\psi_\tau\rangle &=& \frac{1}{\tau}.
\end{eqnarray}
Thus we obtain
\begin{eqnarray}
\tilde J_1 = -\frac{1}{\tau} J_1.
\end{eqnarray}
The same argument implies that $\tilde J_3 = - \frac{1}{\tau}
J_3$.  The essential feature here is that when a FM bond is
decimated the sign of the effective interaction with the
neighboring tau particles is flipped --- FM bonds become AFM bonds
and vice versa.  In addition there is numerical reduction of the
bond strength by a factor of $1/\tau$.  However, as above, the
value of this coefficient is not important for determining the
fixed-point behavior of the model --- the only important fact is
that it is less than 1 so we are guaranteed that $|\tilde J_1| <
|J_2|$.

The two above results for decimation of strong bonds constitute
the strong-randomness RG rules of the mixed random Fibonacci
chain. In the following we will derive the RG flow equations for
this case and show that it has non-trivial fixed points.

\subsection{Flow equations for the Fibonacci chain}

In order to explore the phase diagram of the golden chain, we must
first turn the Ma-Dasgupta rules for the decimation of
ferromagnetic (FM) and antiferromagnetic (AFM) bonds, eq.
(\ref{MD}), to flow equations. This goal was partially achieved in
Ref. \onlinecite{YB} for a golden chain which contains only AFM
bonds. As we shall see, including FM bonds in this analysis
reveals a new fixed point, where the number of FM and AFM bonds is
the same.

We begin our analysis by introducing the logarithmic notation for bond
strengths:
\be
\beta_i=\ln\frac{\Omega}{|J_i|}
\ee
where $\Omega=\max_i\{|J_i|\}$ so that the AFM Ma-Dasgupta rule
(\ref{MD}) reads $\beta_{eff}=\beta_{i-1}+\beta_{i+1}-\ln C$.
Note that while the $\beta_i$'s carry the information about bond strengths, they
do not specify whether a bond is FM and AFM, and $\beta_i\ge 0$. Let
us next define the coupling distributions for the AFM (positive $J$'s)
and FM bonds (negative $J$'s), respectively:
\be
\ba{cc}
P(\beta),  &  N(\beta).
\ea
\ee
The probability of a bond to be AFM (A), or FM (F) are thus
\be
\ba{c}
p_{A}=\intt_0^{\infty}d\beta P(\beta)\\
p_{F}=1-p_{A}=\intt_0^{\infty}d\beta N(\beta). \ea \ee In addition,
we define $\G=\ln \frac{\Omega_I}{\Omega}$ to be the logarithmic
RG flow parameter. Its initial value is a non-universal constant
of order 1, and as the RG progresses, it flows to $\infty$.

The flow equations for $P(\beta)$ and $N(\beta)$ are derived in
analogous fashion to those of the distributions in the spin $1/2$
problem \cite{DSF94}. Roughly speaking, the terms appearing in the
two flow equations are the result of: (a) rescaling of the UV
cutoff $\Omega$, (b) decimation of an AFM bond, (c) decimation of
a FM bond. Below we will write the flow equations with each term
followed by an explanation or a diagram of the decimation step
giving rise to it. In the graphical representation on the right
column below, A and F represent antiferromagnetic and
ferromagnetic bonds respectively, and the bond decimated is
represented by the bold letter with the hat. Let us start with the
flow of the AFM bond distribution:

\begin{widetext}
\be
\ba{cc}
\frac{dP}{d\Gamma}=\hspace{0.5cm}\der{P}{\beta} &  \mbox{(cutoff
  rescaling)}\vspace{3mm}\\
+P(0)\intt_0^{\infty}d\beta_1\intt_0^{\infty}d\beta_2\delta(\beta_1+\beta_2-\beta)P(\beta_1)P(\beta_2)
                          & \bullet\,A\,\bullet\, \widehat{\bf A} \,\bullet\,A\,\bullet\Rightarrow \bullet\,\,A\,\,\bullet\vspace{3mm}\\
+P(0)\intt_0^{\infty}d\beta_1\intt_0^{\infty}d\beta_2\delta(\beta_1+\beta_2-\beta)N(\beta_1)N(\beta_2)
 & \bullet\,F\,\bullet\, \widehat{\bf A} \,\bullet\,F\,\bullet\Rightarrow \bullet\,\,A\,\,\bullet\vspace{3mm}\\
-2P(0) P &  \mbox{(neighbor removal in AFM decimation)}\vspace{3mm}\\
+2N(0) N   & \bullet\,F\,\bullet\, \widehat{\bf F} \,\bullet\,F\,\bullet\Rightarrow \bullet\,\,A\,\,\bullet\,\,A\,\,\bullet\vspace{3mm}\\

-2N(0) P  & \mbox{(removal of neighboring AFM in FM decimation)}\vspace{3mm}\\
+(2P(0)+N(0))P  &
\ea
\ee
The last term feeds back the probability of bonds lost due to an AFM
decimation, which removes a net of two bonds [$2P(0)$], and due to a
FM decimation, which removes a single bond [$N(0)$].

Carrying out the analogous considerations for the FM bond
distribution:
\be
\ba{cc}
\frac{dN}{d\Gamma}=\hspace{0.5cm}\der{N}{\beta} & \mbox{(cutoff
  rescaling)}\\
+2P(0)\intt_0^{\infty}d\beta_1\intt_0^{\infty}d\beta_2\delta(\beta_1+\beta_2-\beta)N(\beta_1)P(\beta_2)
&  \bullet\,F\,\bullet\, \widehat{\bf A} \,\bullet\,A\,\bullet\Rightarrow \bullet\,\,F\,\,\bullet\vspace{3mm}\\
-2P(0) N  & \mbox{(neighbor removal in AFM decimation)}\vspace{3mm}\\
+2N(0) P & \bullet\,A\,\bullet\, \widehat{\bf F} \,\bullet\,A\,\bullet\Rightarrow \bullet\,\,F\,\,\bullet\,\,F\,\,\bullet\vspace{3mm}\\
-2N(0) N  &  \mbox{(removal of neighboring FM in FM decimation)}\vspace{3mm}\\
+(2P(0)+N(0))N.
\ea
\ee
Once more, the last line makes sure that probability is conserved.

Adding up all the above terms yields the following concise flow
equations: \be \ba{c} \frac{dP}{d\Gamma}=\der{P}{\beta}+P(0)\l(P
\otimes P+N\otimes
N)\rr)+2N(0)N(\beta)-N(0)P(\beta)\\
\frac{dN}{d\Gamma}=\der{N}{\beta}+2P(0)N\otimes P-N(0)N(\beta)+2N(0)P(\beta),
\ea
\label{flow}
\ee
where we also introduce the notation:
\be
F\otimes G=\intt_{0}^{\infty} dx_1 \intt_{0}^{\infty} dx_2
\delta(x-x_1-x_2) F(x_1)G(x_2).
\ee
\end{widetext}

\subsection{Fixed points of the real-space RG}

From the flow equations, Eqs. (\ref{flow}), we can find the fixed
points of the golden chain. These appear as attractors of the
integro-differential equations. To find them, we first note that we
can eliminate $\G$ by guessing a scale-invariant solution
\cite{DSF94}:
\be
P_{\G}(\beta)=\frac{1}{\G}p(\beta/\G), \hspace{0.5cm}
N_{\G}(\beta)=\frac{1}{\G}n(\beta/\G).
\ee
Substituting this scaling ansatz gives:
\be
\ba{c}
-p=(1+x)p'+p_0(p\otimes p+n\otimes n)+2n_0 n-n_0 p\\
-n=(1+x)n'+2p_0 n\otimes p +2n_0 p -n_0 n.
\label{flow1}
\ea
\ee
Furthermore, the convolution hidden by the $\otimes$ sign compells us
to assume an exponential form for the unknown functions $n(x),\,p(x)$:
\be
\ba{cc}
p(x)=p_0 e^{-\gamma x} & n(x)=n_0 e^{-\gamma x}.
\ea
\ee
This ansatz reduces the integro-differential equations,
Eqs. (\ref{flow}), to a set of three simple algebraic equations:
\be
\ba{c}
\gamma=p_0^2+n_0^2\\
n_0(\gamma-2p_0^2)=0\\
n_0+p_0=\gamma.
\ea
\label{FPeq}
\ee

The exponential ansatz and the resulting Eqs. (\ref{FPeq}) reveal two fixed-point solutions.
A first solution of (\ref{FPeq}) corresponds to the pure AFM fixed
point:
\be
\ba{cc}
n_0=0. & p_0=\gamma=1.
\ea
\ee
This is the random singlet phase discussed in
Ref. \onlinecite{YB}. A new fixed point, however, is found
by allowing $n_0$ to be nonzero:
\be
\ba{cc}
\gamma=2, & p_0=n_0=1.
\ea
\ee
This fixed point has an equal proportion of FM and AFM bonds, and
although it is an infinite-randomness fixed point, it is not a
random-singlet point. Translating back to the original variables, the
coupling distributions are:
\be
N(\beta)=P(\beta)=\frac{1}{\Gamma}e^{-2\beta/\Gamma}.
\ee

While $\gamma=2$ is one universal critical exponent describing the
universality class of the mixed FM/AFM phase, another critical
exponent is $\psi$, which describes the energy-length scaling: \be
\ln\frac{1}{E}\sim L^{\psi}. \ee This is equivalent to: \be n\sim
\frac{1}{\Gamma^{1/\psi}}, \ee with $n\sim 1/L$ here being the
density of undecimated sites.

To obtain $\psi$, let us compute the density of free sites at the energy scale
$\G$. A FM bond decimation eliminates one site, while an
AFM decimation eliminates two sites. This implies that the total
density of undecimated sites obeys:
\be
\frac{1}{n}\frac{dn}{d\G}=-N(0)-2P(0)=-3\frac{1}{\G}
\ee
and therefore:
\be
n=\frac{n_I}{\Gamma^3},
\ee
which corresponds to the infinite randomness critical exponent:
\be
\psi=1/3.
\ee

From $\psi$ and $\gamma$ of the mixed FM/AFM fixed point of the
Fibonacci random chain, we see that it is in the same universality
class as the fixed point separating the gapped Haldane-phase and
the random singlet phase of the $S=1$ random chain.

\subsection{Stability of the phases}

In the previous section we found the two fixed points of the random
golden chain. In order to construct its phase diagram, however, we
must also study the stability of these fixed points. As it turns out,
the random-singlet phase is actually {\it unstable}, and flows to the
mixed FM/AFM fixed point.

Let us begin our analysis with the mixed FM/AFM phase. Assume a
perturbation that breaks the balance between FM and AFM bonds:
\be
\ba{cc}
N(\beta)=(1-\delta)\frac{1}{\G}e^{-2\beta/\G}, & P(\beta)=(1+\delta)\frac{1}{\G}e^{-2\beta/\G}.
\ea
\ee
Substituting in Eqs. (\ref{flow}) very readily yields:
\be
\G\frac{d\delta}{d\G}=-5\delta,
\ee
indicating stability with respect to FM/AFM imbalance.

\begin{figure}
\includegraphics[width=9cm]{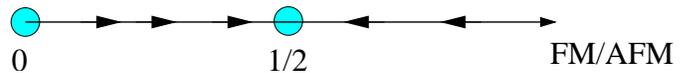}
\caption{Flow diagram for the mixed Fibonacci chain. Two fixed points
  exist. The AFM fixed point is unstable with respect to introduction
  of FM bonds. A stable fixed point exists at the symmetric FM-AFM
  point. \label{taufig2}}
\end{figure}

By establishing the stability of the mixed phase, we essentially
doom the random singlet phase to be unstable. Complementing the
analysis, however, near the AFM random singlet fixed point, we
assume: \be \ba{cc} N(\beta)=\delta\frac{1}{\G}e^{-\beta/\G} &
P(\beta)=(1-\delta)\frac{1}{\G}e^{-\beta/\G}. \ea \ee Again,
substitution in Eqs. (\ref{flow}) yield: \be
\G \frac{d\delta}{d\G}=2\delta, \ee which means that FM
bonds are a relevant perturbation. In addition, we find that the
cross-over exponent is: \be \chi=2. \ee These results allow us to
draw the flow diagram, Fig. \ref{taufig2}. Thus the golden chain
is in the AFM random-singlet phase when it initially consists of
only AFM bonds. On the other hand, any finite density of FM bonds
leads to the mixed FM/AFM fixed point.

\section{Entanglement entropy at the symmetric FM-AFM point \label{ee}}

\subsection{Overview of the calculation}

\begin{figure}
\includegraphics[width=6.5cm]{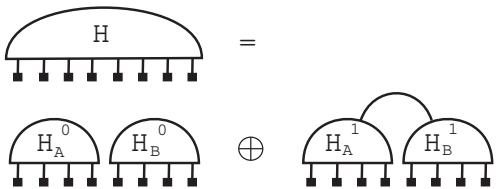}
\caption{Decomposition of the total Hilbert space into superselection sectors. \label{ff11}}
\end{figure}

In this section we calculate the asymptotic scaling of the block
entanglement entropy of the disordered Fibonacci chain, that is,
the entanglement entropy between a region of $L$ consecutive sites
and its complement. Because of the non-local nature of the Hilbert
space, some subtleties arise. Let us first define entanglement
entropy, and then motivate our definition. Given two regions $A$
and $B$, we have, as illustrated in fig. \ref{ff11},
superselection sectors for the topological charge, with the total
Hilbert space \be \label{Hdecomp} H = {H_A}^0 \otimes {H_B}^0 \,
\oplus \, {H_A}^1 \otimes {H_B}^1. \ee Here the superselection
sectors ${H_A}^i$ and ${H_B}^i$ can formally be thought of as
$n$-point disk spaces. Given a state $\psi \in H$ we decompose it
according to (\ref{Hdecomp}) as $\psi = \psi^0 + \psi^1$.  Each of
these has a Schmidt decomposition \be \psi^i = \sum_j
{\lambda_j}^i {\eta_j}^i \otimes {\chi_j}^i \ee where the states
${\eta_j}^i \otimes {\chi_j}^i$ have unit norm in $H$. We now
define the entanglement entropy in the usual way, as \be S = -
\summ_{i,\,j} {\lambda_j}^i \log_2 {\lambda_j}^i. \ee

To motivate this definition we note that it is equivalent to the
standard definition of entanglement entropy when we implement the
fusion rule constraints via large energy penalties $E$ in the
Hamiltonian.  Specifically, working in the link basis for convenience,
we enlarge the Hilbert space to a space $H'$ that allows all link
configurations, with terms added to the Hamiltonian to penalize
violations of the fusion rules.  We extend the inner product and
Hamiltonian to $H'$ in the simplest way possible - say, by extending
the Hamiltonian to be $k E$ times the identity on the space $V_k$ of
configurations with $k$ violations of the fusion rules, and taking the
inner product such that $V_k$ is orthogonal to $V_l$ for $k \neq l$.
The new hamiltonian is Hermitian on $H'$, preserves $H \subset H'$,
and reduces to the original Hamiltonian on $H$.  For $E$ much larger
than the ground state energy in the original problem, the ground state
and all low lying states in the new problem are the same as those in
the original one.  The upshot is that $H'$ now has a tensor product
decomposition, and entanglement entropy can be defined in the
conventional way (some care must be taken in normalizing inner
products on the sub-system Hilbert spaces).  This conventional
definition for states in $H \subset H'$ coincides with ours above.  We
also note that this is how entanglement entropy was defined in the
numerical algorithm of Ref. \onlinecite{Qpeople1}, which recovered the central
charges $c=4/5, 7/10$ in the uniform case of the golden chain.

The calculation of the entanglement entropy for the Fibonacci
chain proceeds along the same lines as previous calculations of
the entropy. To gain orientation for the calculation we will
shortly pursue, let us review the entanglement entropy calculation
for the simplest instance of an infinite randomness fixed point,
the spin $1/2$ Heisenberg model \cite{RefaelMoore2004}.  There,
the basic idea is to count the number of singlets formed over a
boundary of the interval, up to a cutoff size $L$ (energy-length
scaling turns this into a cutoff in $\Gamma$).  Each singlet
contributes $1$ to the entanglement entropy.  Real space RG
analysis shows that the number of singlets is proportional to
$\log \Gamma \sim \log L$, so we obtain logarithmic scaling of the
block entanglement entropy.  It turns out that this logarithmic
scaling persists in the disordered fibonacci chain, but obtaining
the coefficient in front of the $\log$ is considerably more
difficult.  For one thing, we already saw earlier (discussion
preceding fig. \ref{ff5}) that even in the AFM fixed point,
obtaining the entanglement entropy required using $F$-moves to
change to a more convenient basis.

Obtaining the entropy in the mixed fixed point of the Fibonacci
chain is even more difficult, because the ground state now
contains not just singlets but also complicated tree-like
structures, since two $\tau$'s can fuse into another $\tau$, and
not just to a singlet.  The problem, however, is still tractable,
although instead of looking at the RG time between successive
singlets, we must now look at RG times between consecutive AFM
decimations, and the tree-like structures that form between them
(fig. \ref{ff6}a).  Just as each singlet in the AFM case
contributed asymptotically $\log_2 \tau$ in the case of many
singlets, we will find a similar simplification in the mixed case
for a large number of $AFM$ decimations - each tree-like structure
will asymptotically contribute some amount to the entropy.  The RG
process will average over all trees, so we will have some average
contribution $<S_{\text{tree}}>$ to the entanglement entropy.  To
get the dependence on the block length $L$, we first use
energy-length scaling to relate $L$ to the RG flow variables:
$\frac{1}{3} \ln L \sim - \ln \Gamma \sim l$.  We show in the next
subsection that AFM decimations occur with period $3/2$ in $l$.
Thus for a block of size $L$, we have $n \sim \frac{2}{3} l \sim
\frac{2}{9} \ln L$ AFM decimations separating tree-like structures
straddling each endpoint of the block, which gives a contribution
of

\begin{equation}\label{stree}
\frac{4}{9} \, <S_{\text{tree}}> \, \ln L.
\end{equation}

We will see that there will also be another contribution to the entanglement entropy, coming from the residual singlets left straddling the endpoints after the tree-like structures have been resolved.  We will compute this "rainbow" contribution carefully later in this section, but first we turn to calculating the RG times between the various decimations.

\subsection{RG times between decimations}

\begin{figure}
\includegraphics[width=8cm]{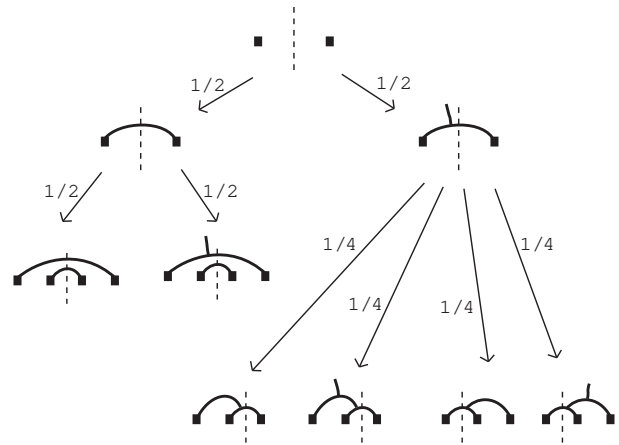}
\caption{RG history near the partition bond (represented by the dashed
  vertical line).  The partition bond is at some point decimated, with
  equal probabilities of having an AFM and FM decimation.  In the case
  of an AFM decimation, we get another partition bond, which will at
  some point be decimated by an AFM or FM decimation, again with equal
  probability.  In the case of an FM decimation, however, we produce a
  partition site, which can decimate, either via AFM or FM decimation,
  with either the site to its left or the site to its right.  All four
  of these possibilities are equally likely.  Carried out further,
  this process will generate complicated tree-like structures.  Note
  that we have ignored decimations that do not involve the partition
  bond or partition site \label{ff12}}
\end{figure}

A pictorial representation of the RG process is given in fig.
\ref{ff12}.  We see that eventually a ground state of the form
shown in fig. \ref{ff6} is generated.  To quantitatively
understand real space RG, we will need to compute the ($\log$) of
the RG times between various types of decimations.

Now, in the simple case of the spin $1/2$ Heisenberg model, one is
interested in ($\log$ of) the RG time between successive
decimations of the partition bond - the bond through which the
boundary of the region passes.  After each decimation of the
partition bond, the coupling distribution at the bond is different
than the average distribution in the chain. Nevertheless, it is
independent of the other surviving bonds \cite{RefaelMoore2004},
and in that sense, it 'resets'. Thus the RG time duration between
successive decimations obeys a Poisson distribution characterized
by one number, the average of the $\log$ of the RG time between
successive decimations. No history dependence of this number
appears for the antiferromagnet.

Our mixed Fibonacci case is more complicated due to the presence
of both FM and AFM decimations.  Here we must consider all
possible histories of FM and AFM decimations, as illustrated in
fig. \ref{ff12}, and compute probabilities for each. Nevertheless,
the Fibonacci chain has several simplifying factors which make the
problem more tractable. First, as in the case of the spin-1
Heiseneberg chain calculation, once the partition bond undergoes
an AFM decimation, the resulting distribution of the coupling
across the partition bond is 'reset', and is independent of the
chain's history. Second, the distribution of bond strengths is
always symmetric with respect to exchange of FM and AFM couplings.
It turns out that it is characterized by just two numbers, the
time between an AFM decimation and the next decimation (equally
likely to be FM and AFM by symmetry), and the time between a FM
decimation and the next one (again equally likely to be FM and
AFM).

To see this, we notice that the joint probability distribution of all
the bonds takes one of two forms, depending on whether we've just had
a FM or AFM decimation.  Immediately following an AFM decimation, we
have an independent distribution for all bonds, with the partition
bond having distribution
\be
Q(\beta) = \frac{2 \beta} {\Gamma^2} \, e^{-2 \beta / \Gamma}
\ee
and all the other bonds following $P(\beta) = \frac{1}{\Gamma} \,
e^{-2 \beta / \Gamma}$.  FM decimations are even simpler: after an FM
decimation, all the bonds follow $P(\beta) = \frac{1}{\Gamma} \, e^{-2
  \beta / \Gamma}$.  The surrounding bonds do get changed from AFM to
FM and vice versa, but because they are equally likely to be one or
the other at the mixed fixed point (i.e., the distribution is symmetric with respect to the
interchange) there is no net effect.  One can verify these
observations by noting, as in \onlinecite{RefaelMoore2004}, that under RG
evolution following an AFM decimation the distribution retains its
form, with $Q$ changing in a complicated way and $P$ evolving as its
explicit dependence on $\Gamma$ dictates.  In fact, in much the same
way as is done in Ref. \onlinecite{RefaelMoore2004} we can derive an
equation for the RG evolution of $Q$:
\begin{eqnarray}
\label{evolQ} && \frac{d Q_{\Gamma} (\beta)}{d \Gamma} =
{Q'}_{\Gamma} (\beta) - \frac{2}{\Gamma}Q_{\Gamma}(\beta) + \\ &&
\frac{4}{\Gamma} \int d\beta_1\, d\beta_2 \, \delta(\beta -
\beta_1-\beta_2) \, P_{\Gamma} (\beta_1) \, Q_{\Gamma} (\beta_2).
\nonumber
\end{eqnarray}
We solve it by making the ansatz
\be
\Qg= \left(a + b \frac{2
  \beta}{\Gamma} \right) \frac{1}{\Gamma} \, e^{-2 \beta / \Gamma},
\ee
with $a$ and $b$ functions of $\Gamma$.  Let $l = \ln
\frac{\Gamma}{\Gamma_0}$, where $\Gamma_0$ is the RG time when the AFM
decimation occurred.  The initial conditions at $l=0$ are then $a=0,
b=1$.  Plugging the ansatz into (\ref{evolQ}) then yields
\begin{eqnarray}
 \frac{da}{dl}= -3a + 2b, \nonumber \\
\frac{db}{dl}= a - 2b.
\end{eqnarray}
The solution is
\begin{eqnarray} a=\frac{2}{3}
  \, (e^{-l} - e^{-4l}), \nonumber \\
 b=\frac{1}{3} \, (2 e^{-l} + e^{-4l}).
 \end{eqnarray}
To extract the expected value $<l>$ until
the next decimation, note that the probability $p$ that another
decimation has not occurred by RG time $\Gamma > \Gamma_0$ is simply
\be
p_{\Gamma} = \int_{0}^{\infty} \, d\beta \, Q_{\Gamma} (\beta) =
\ag + \bg.
\ee
The expected value of $l$ at the next decimation is
then
\begin{eqnarray}
<l> &=& - \int_{0}^{\infty} l \, d p_{\Gamma} =
  2 \int_{0}^{\infty} a \,l\, dl \nonumber \\ &=& \frac{4}{3}
  \int_{0}^{\infty} (e^{-l} - e^{-4l})\,dl = \frac{15}{12}.
\end{eqnarray}
Notice that $<l>$ is independent of $\Gamma_0$.

In a similar manner, we can consider the case where an FM
decimation of the partition bond has just occurred at $\Gamma_0$.
In this case all the bond strengths just evolve according to
independent distributions $P(\beta) = \frac{1}{\Gamma} \, e^{-2
\beta / \Gamma}$, so the whole situation is characterized by an
overall probabilistic weight $w$.  The RG equation for $w$ is
readily derived to be \be dw\,=-4\,w\,\frac{d\Gamma}{\Gamma}, \ee
where the prefactor of 4 is due to the fact that the site
containing the partition can be decimated by processes on either
side of it which can each be either FM or AFM, thus leading to
four possibilities. This equation  is solved by $w=(\Gamma /
\Gamma_0)^{-4} = e^{-4l}$. The expected value of $l$ at the next
decimation is then \be <l> = - \int (d e^{-4l})\, l = 4 \int
e^{-4l} \, l \, dl = \frac{1}{4}. \ee

Finally, we compute the average $<l>$ between AFM decimations, as
follows.  First, note that there can be any number of FM decimations
in between the two AFM decimations.  Since each decimation is equally
likely to be AFM and FM, the probability of having precisely $k$  FM
decimations is is $2^{-k-1}$.  The expected $l$ is therefore
\begin{eqnarray} \label{expl} <l> = \sum_{n=0}^{\infty} \left(\frac{15}{12} +
  \frac{1}{4} n \right) 2^{-n-1} = \frac{15}{12} + \frac{1}{4} =
  \frac{3}{2}. \label{ft} \end{eqnarray}

\subsection{Entropy Calculation}

We now use the knowledge of mean decimation times to do the entropy
calculation.  We are interested in the scaling limit of large $L$,
which translates to looking at large $l = \log
\frac{\Gamma}{\Gamma_0}$.  In this case a complicated tree-like
structure forms over each endpoint of the length $L$ interval, and we
need to figure out its entropy contribution.  As we mentioned above,
for the case of the random Fibonacci chain with only $AFM$ couplings,
where the picture is a rainbow diagram straddling each endpoint, the
asymptotic contribution in the large rainbow limit of each singlet in
the rainbow is $\log_2 \tau$.  We will now find an analogous rainbow
picture for the mixed Fibonacci chain.

\begin{figure}
\includegraphics[width=5cm]{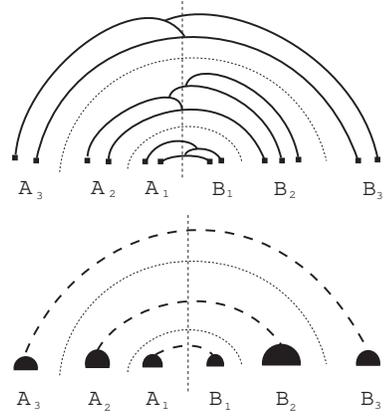}
\caption{a) Trivalent graph representing a ground state generated
by the real space RG.  It can be decomposed into tree components,
which are separated by AFM decimations of the partition bond,
denoted in the figure by dotted lines.  There could also be tree
diagrams that don't straddle the partition bond and hence do not
contribute to the entanglement entropy; we omit them from the
illustration for clarity.  b) After applying $F$-matrix relations
we can reduce the trivalent graph in a) to a superposition of
graphs of this form.  Here the dashed lines denote either a $1$
(trivial) or $\tau$ (nontrivial) line.
  \label{ff6}}
\end{figure}

Let us focus on just one boundary of the interval, so we have one
partition bond.  The ground state trivalent graph that forms over it
can be decomposed into connected tree components (fig. \ref{ff6}a) as
follows.  The first, innermost tree is generated by all the FM
decimations prior to the first AFM decimation of the partition bond
(if the first decimation is AFM, the tree is just a singlet).  The
next tree is generated by all the FM decimations between the first and
second AFM decimations, and so on.  As shown in (fig. \ref{ff6}a)
these trees can be thought of as "thickened stripes" - it's just that
now the stripes consist of not only a singlet, but an entire tree
straddling the partition bond.  The idea now is to use $F$-matrix
relations to decompose each tree into a superposition of graphs which
have only $0$ or $1$ $\tau$ lines straddling the partition bond, so as
to get the ground state to look like a superposition of rainbow
diagrams (fig. \ref{ff6}b).  The entropy will then be a sum of a
rainbow contribution and a contribution coming from the entanglement
between the graphs on either side of the partition bond joined by the
rainbow stripes.

\begin{figure}
\includegraphics[width=8cm]{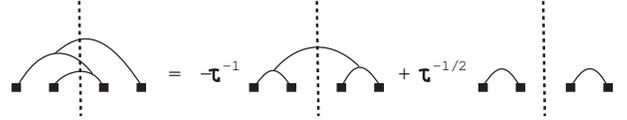}
\caption{Application of one $F$-matrix move decomposes the graph on the left into the superposition on the right. \label{ff9}}
\end{figure}

Before we go into details, let's compute a specific example.
Consider the graph in fig. \ref{ff9}, which describes two FM
decimations followed by an AFM one.  To compute its entanglement
entropy we apply the $F$-matrix move as shown in fig. \ref{ff9}
and decompose it into a superposition of the two graphs on the
right side of the equation.  The entanglement entropy is then
$-\tau^{-2} \log_2 \left( \tau^{-2} \right) - \tau^{-1} \log_2
\left( \tau^{-1} \right)$.  In general we'll need to apply many
$F$-matrix moves and the superposition will be more complicated.

Let's proceed step by step.  Label the ``thickened stripes"
(trees) in fig. \ref{ff6}(a) by an index $i$, running from $1$ to
the number of thickened stripes $n$, and suppose the $i$th stripe
connects a region $A_i$ on the left side of the partition to a
region $B_i$ on the right side.  Consider Hilbert spaces $H_{A_i}$
and $H_{B_i}$ associated with these sites.  These are spanned by
trivalent graphs having endpoints on those sites, as before, but
this time, because the regions may have nontrivial topological
charge, we have the familiar decompositions $H_{A_i} = {H_{A_i}}^0
\, \oplus \, {H_{A_i}}^1$ and $H_{B_i} = {H_{B_i}}^0 \, \oplus \,
{H_{B_i}}^1$.  The index $0$ and $1$ just corresponds to whether
or not $A_i$ and $B_i$ are connected by a $\tau$ line.  The
Hilbert Space $H_i$ of the union is \be H_i = {H_{A_i}}^0 \otimes
{H_{B_i}}^0 \, \oplus \, {H_{A_i}}^1 \otimes {H_{B_i}}^1. \ee  The
ground state is a product state in $H = \bigotimes_i H_i$.  Using
the decomposition for $H_i$ we write the factors  $\psi_i =
\alpha_i {\psi_i}^0 + \beta_i {\psi_i}^1$.  Here ${\psi_i}^j$ are
normalized to have norm $1$ and $|\alpha_i|^2 + |\beta_i|^2 = 1$.
For convenience we take $\alpha_i$ and $\beta_i$ real and
positive.  Let $\gamma$ be the average over $i$ of $|\beta_i|^2$.
When we foil the above product we get the ground state as a
superposition of $2^n$ states with differing rainbow
configurations. One can check (by taking inner products and using
the no tadpoles rule) that not only are all these states
orthogonal, but all the states that enter into the Schmidt
decomposition of one (on, say, the left side) are orthogonal to
all the states that enter the Schmidt decomposition of the other.
This yields a block diagonal decomposition of the density matrix
of, say, the left side.  Thus we can deal with the blocks in this
block diagonal decomposition separately. We label these $2^n$
blocks with a label $b$.  Each block $b$ corresponds to a choice
of $h_i = 0,1$, where $h_i$ specifies the topological charge of
$A_i$.  The trace of such a block $b$ is \be t_b = \prod_i
{\alpha_i}^{2h_i} {\beta_i}^{2(1-h_i)}. \ee

\begin{figure}
\includegraphics[width=6cm]{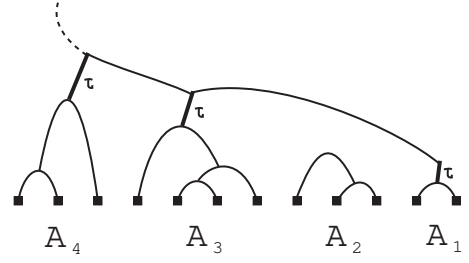}
\caption{A trivalent graph whose labelings give a basis for the
Hilbert space left of the partition bond.  Here we have a tree for
each region $A_i$, with the line leading out of $A_i$ fixed to be
$\tau$ (for $i=1,3,4$) or $1$ (for $i=2$).  The rest of the lines,
including the lines connecting different $A_i$, can be labeled at
will, consistent with the fusion rules.  The label of the dashed
line gives the total topological charge of the left side of the
system.\label{ff7}}
\end{figure}

Let's compute $-\text{Tr} \, \left( M_b \log M_b \right)$ for
block $b$.  To do this we choose a convenient basis for the space
where the, say, left components of the Schmidt decomposition of
the state corresponding to this graph lie.  In general, a basis
can be given by the set of all labelings of a trivalent tree,
consistent with the fusion rules (i.e., you can't have two $1$'s
and a $\tau$ at a vertex). So we pick a tree for each region
$A_i$, and then join these up as in Fig. \ref{ff7}, for a tree
defined over the whole left region (note that we're in the
subspace where the topological charges of each $A_i$ are fixed by
$h_i$, so we're not looking at {\it all }labelings, but fixing
some of the edges to be $\tau$). This is precisely the kind of
graph used in \onlinecite{YB} to compute the entropy of the
rainbow diagram. The only difference in our case is that we have
extra degrees of freedom corresponding to the graphs for each
$A_i$.  So we've found a basis which consists of labelings of
several subgraphs of a trivalent tree which do not interfere with
each other.  Namely, these subgraphs are the graphs near each
$A_i$, and an extra one consisting of the edges which link the
various $A_i$ - it will lead to the ``rainbow" contribution in the
equation below. Because there are no {\em inter}-sub-graph
constraints, i.e., the labelings on the sub-graphs can be chosen
independently, the entropy is simply a sum of contributions from
each subgraph. \be -\text{Tr} \, \left( M_b \log M_b \right) =
-t_b \, \log t_b + t_b \, {S_b}^{\text{rainbow}} + t_b \, \sum_i
{S_i}^{h_i} \ee Here the first term comes from the normalization
of the block $b$, which has overall trace $t_b$, the second term
is the ``rainbow" contribution mentioned above, asymptotically
equal to $\log_2 \tau$ times the number of nonzero $h_i$, and
${S_i}^j$ is the entropy associated to ${\psi_i}^j$ (i.e the
contribution from the graphs around $A_i$ and $B_i$.  Summing over
all blocks $b$ and performing some elementary algebra, we get the
entropy to be
\begin{eqnarray} S&=&{\sum_i} \left( -2 {\alpha_i}^2 \log \alpha_i - 2 {\beta_i}^2 \log \beta_i + {\alpha_i}^2 {S_i}^0 + {\beta_i}^2 {S_i}^1 \right) \nonumber \\ &+& <{S_b}^{\text{rainbow}}> \end{eqnarray} where the average $<{S_b}^{\text{rainbow}}>$ is taken over all blocks $b$ with weight $t_b$.  This average is approximately equal to $\log_2 \tau$ times the average number of stripes, $\gamma n$, so that
\begin{eqnarray} \label{entf} S &\sim& {\sum_i} \left( -2 {\alpha_i}^2 \log
  \alpha_i - 2 {\beta_i}^2 \log \beta_i + {\alpha_i}^2 {S_i}^0 + {\beta_i}^2 {S_i}^1
  \right) \nonumber \\ &+& \gamma n \log_2 \tau
\end{eqnarray}

Recall that we defined $\gamma$ above to be the average over $i$
of ${|\beta_i|}^2$, which is just the fraction of regions $A_i$
which have non-trivial topological charge.

In eqn. \ref{entf} the first quantity sums up the tree
contributions $S_{\text{tree}}$ discussed at the beginning of this
section.  Thus eqn. \ref{stree} shows that it is equal to
\begin{equation} \frac{4}{9} <S_{\text{tree}}> \ln L.
\end{equation} To compute $<S_{\text{tree}}>$ we note that the average is
taken over all possible trees generated by FM decimations between
two consecutive AFM decimations.  There are many possible trees,
since we can choose the number $r$ of FM decimations, and for each
FM decimation we must decide whether to decimate with the right or
the left site.  The probability of each such tree is $2^{-2r-1}$.
We have computed this average numerically via a Mathematica
program.  The program basically takes each possible tree and
builds up the corresponding wavefunction $\psi$ step by step in a
convenient basis by applying the FM decimations.  It then traces
out half the system, finds the eigenvalues of the density matrix,
computes the entanglement entropy, and finally averages over the
trees.  We obtain $<S_{\text{tree}}> = 0.115$ approximately.  The
program also computes $\gamma = 0.927$.  Putting this into
(\ref{entf}) we get $S=\frac{4}{9} \, \ln L \, \left(0.115 + 0.927
\, \log_2 (\tau)\right)$ so that \be \label{entfinal} S =  0.234
\, \log_2 L. \ee

The program goes up to $r=9$, and we can bound the
error obtained by omitting the remaining trees by a quantity
exponentially small in $r$.  Basically this is because the probability
of having a tree with a given value of $r$ is exponentially small,
$2^{r+1}$, whereas the maximal entropy contribution of such a tree
only scales linearly in $n$ (because the dimension of the Hilbert
space is exponential in $n$).  This argument yields a rigorous bound
of $\pm 0.0006$ on the coefficient in (\ref{entfinal}).

Thus the effective central charge we obtain for the mixed fixed point
is:
\be
c_{eff}^{mixed}\approx 3\cdot 0.234 = 0.702.
\ee
As we will discuss in the conclusion, this result is bigger than the
effective charge in the antiferromagnetic random-singlet fixed point.

\begin{figure}
\includegraphics[width=5cm]{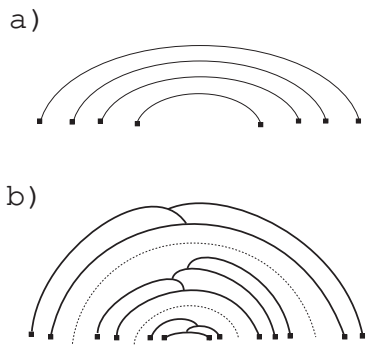}
\caption{a) A rainbow diagram.  b) Typical state at the FM-AFM point.  The dashed lines denote the location of AFM decimations, which separate the ``thickened" stripes. \label{ff1}}
\end{figure}

\section{Non-abelian anyons and infinite disorder}

\subsection{Fusion rules and real-space RG}

In this paper so far we analyzed the infinite disorder fixed points of the
Fibonacci model. Rather surprisingly, the Fibonacci anyons lend
themselves very readily to real-space RG analysis, and gives rise to a
remarkably rich phase diagram. It is then natural to ask: will similar
effects arise in other chains of non-abelian anyons?

In fact, a deep relationship exists between real space RG and the
behavior of non-abelian anyons. To see this consider first the
ground state of conventional spin chains. In order to find the
ground state of a conventional spin chain in a strong-disorder
phase, we would apply real-space decimation rules to bond with
strong coupling. The type of decimation we apply will depend on
the local Hamiltonian and the Hilbert space of the system. For
instance, in the spin-1/2 (AFM) Heisenberg model, two neighboring
spin-1/2's can fuse according to the the SU(2) rule: \be
\frac{1}{2}\otimes\frac{1}{2}=0\oplus 1. \label{FRhalf} \ee A
decimation rule applied to these two neighboring spins will choose
one of the fusion subspaces - the spin-singlet or spin-triplet -
according to the local Hamiltonian. The generalization of this
principal to the case of non-abelian chains is nearly trivial. The
spin-compounding rule, Eq. (\ref{FRhalf}), is substituted by the
{\it fusion algebra} of the non-abelian system: \be a\otimes b
=\oplus\summ_c N^{c}_{ab} c \label{FA} \ee where $N_{ab}^{c}$ is
the number of ways the superselection sectors $a$ and $b$ can fuse
into $c$.

A major difference, however, between rules (\ref{FRhalf}) and
(\ref{FA}) is that fusion rules for a non-abelian algebra are always {\it
  closed}, while in regular spin-chains, the fusion rules include an
infinite set of subspaces. The closure of the fusion rules for
non-abelian anyons is a manifestation of the nonlocality of their
Hilbert space, and therefore unique to these systems. It implies
that one can {\it always} apply a real-space RG scheme without
ever generating new types of coupling in the renormalized
Hamiltonian. Furthermore, just as in conventional spin chains, a
decimation will result in either in a Ma-Dasgupta renormalization
of the neighboring couplings, or in their multiplication by a
factor of magnitude smaller than 1. Therefore {\it sufficiently
disordered (and most likely even weakly disordered) non-abelian
chains will exhibit an infinite randomness behavior in the large
length scale properties of their ground state.}

\subsection{$S>1/2$ Heisenberg chains and the $SU(2)_k$ fusion
  algebra}

The above observation is easily demonstrated using the mixed
FM/AFM fixed point of the Fibonacci anyons. Both FM and AFM
couplings between two Fibonacci anyons lead to fusion into either
a Fibonacci anyon, or the vaccum: \be \tau\otimes \tau ={\bf
1}\oplus\tau. \label{tauFA} \ee Therefore we can generically carry
out a real-space RG analysis to its conclusion. But in spin-1/2
chains with nearest-neighbor couplings that could be either FM or
AFM, it is easy to see that we generate higher and higher spins,
and as a result do not flow to an infinite randomness fixed point
(although a fixed point of the mixed spin-1/2 chain was observed
numerically in \onlinecite{LeeSigrist1,LeeSigrist2}), unless their
Hamiltonian is restricted, e.g., by a symmetry in the problem
which prevents large-moments formation. This is the case in
$S>1/2$ Hiesenberg models \cite{MGJ,HymanYang, RKF}, which we will
now briefly discuss.

Disordered Heisenberg spin chains with spin $S>1/2$ were
successfully analyzed by a real space decimation procedure that
instead of forcing two strongly interacting sites into their
lowest energy subspace (usually the singlet), just forbids them
from their highest energy subspace, (usually with spin $2S$). This
gives rise to sites becoming effectively lower-spin sites, with
spins $S_i=1/2,\ldots,S$ \cite{MGJ,
  HymanYang}. Although
the bare Hamiltonian contains only antiferromagnetic couplings,
the decimation procedure also generates ferromagnetic bonds. These
raise the spectre of large-spin moment formation, but the
bipartiteness of the chain in the bare Hamiltonian guarantees that
these FM coupling can never give rise to a spin larger than the
original spin.

The fact that in disordered Heisenberg models the real-space
decimation rules only allow the formation of spins no larger than
the original spins, makes these rules  almost identical to the
fusion rules of the truncated $SU(2)$ representations, $SU(2)_k$
with $k=2S$. As an example consider the spin-1 Heisenberg model.
After some renormalization, the spin-1 chain effectively contains
sites completely decimated, spin-0, partially decimated, spin-1/2,
and sites that are spin-1. Upon reals-space decimation, the ground
state is formed by the following 'fusion' rules: \be \ba{c}
1\otimes 1=0\\
1\otimes 1/2=1/2\\
1/2\otimes 1/2=0\oplus 1 \ea \label{s11} \ee These fusion channels
are picked {\it energetically}; i.e., two spin-1 sites can fuse
into a spin-2 moment, but this will be very costly, and will be
excluded from the ground state wave function. The two fusion
possibilities of the spin-1/2 in the last line indicate that
spin-1/2's can have FM and AFM interactions. If we now compare
this to the $SU_2(2)$: \be \ba{c}
\epsilon\otimes\epsilon={\bf 1}\\
\epsilon \otimes \sigma =\sigma\\
\sigma\otimes\sigma={\bf 1} \oplus \epsilon \ea \label{su22} \ee
we can identify the non-trivial superselection sector, $\sigma$,
with spin-1/2, as expected form the Bratteli diagram, and the
trivial sector $\epsilon$ with spin-1.

Indeed the two fusion rules are essentially identical. But as
opposed to rules (\ref{s11}), which are imposed by energy
consideration, the fusion rules (\ref{su22}) are complete, and
describe the full Hilbert state, rather than the ground state.
Therefore the ground state of a disordered $SU_2(2)$ chain is
different than that of a spin-1 disordered Heisenberg chain. In
fact, the $SU_2(2)$ reduces to a random Majorana chain, analyzed
in Ref. \onlinecite{YB}, which, has a random singlet ground state.
A similar situation prevails in the case of the spin-3/2
Heisenberg model \cite{RKF}: the decimation rules for the spin
model are almost exactly the same as the fusion rules for
$SU_3(2)$ (except for the Heisenberg model {\it not} allowing the
fusion $1\otimes 1=1$, which could be corrected by allowing
biquadratic coupling). Nevertheless, the spin-3/2 Heisenberg chain
exhibits two random singlet phases, separated by a 4-domain
permutation symmetric fixed point, while the behavior of the
non-trivial sector of the $SU_3(2)$  is given by the above
analysis of the Fibonacci chain.

\subsection{Novel infinite randomness universality classes in
  non-abelian anyons?}

Although the analogy between the spin $S>1/2$ Heisenberg model
decimation rules, and the fusion rules of $SU(2)_k$ theories does not
help us find new ground states, it demonstrates something rather
important. Just as the random Heisenberg models allowed the discovery
of the permutation symmetric sequence of infinite randomness fixed
points \cite{DamleHuse}, we expect that an investigation of disordered
$SU(2)_k$ chains will also lead to novel infinite randomness
universality classes. We leave this study for future research.

\subsection{Limit of $k\rightarrow \infty$}

In the case of the disordered AFM golden chain, it is possible to
generalize the setup slightly by considering other larger values
of $k$.  Retaining the quantum spin $1$ representation for the
anyons out of which we build the chain, the entire analysis of
\onlinecite{YB} goes through, with only the quantum dimension
changed from $\tau$ to $2 \cos (\pi / (k+2))$.  It is gratifying
to see that in the ``classical" limit $k \rightarrow \infty$, we
reproduce the spin $1/2$ result, with each singlet contributing
$\log_2 2 = 1$ to the entanglement entropy.

\section{Conclusions}

In this paper, we carry out an exhaustive analysis of the simplest
random chain of non-abelian quasiparticles: the Fibonacci, or
golden, chain. Using real-space RG, we are able to analyze the
phase diagram and stability of the entire parameter range of the
nearest-neighbor Fibonacci chain, where each pair of neighboring
sites interacts by assigning an energy cost for fusing in the
trivial channel or in the anyonic channel.

The phase diagram we find is split between two phases, both of
which are infinite randomness phases.  When there are only
couplings favoring fusion into the trivial channel (i.e., only AFM
couplings) the flow is to the random singlet fixed point. When any
finite density of 'ferromagnetic couplings', i.e., couplings
preferring the $\tau$ fusion channel, are sprinkled in, the random
singlet fixed point is destabilized, and the chain flows to a
mixed infinite-randomness phase, which is characterized by the
energy-length scaling exponent $\psi=\frac{1}{3}$, and a coupling
distribution function
$\rho(J)\propto\frac{1}{|J|^{1-\chi/\Gamma}}$, with $\chi=2$, for
both FM and AFM cite. This stable fixed point, somewhat
surprisingly, is in the same universality class as the transition
point between the Haldane phase and the random singlet phase of
the random spin-1 Heisenberg model \cite{HymanYang, MGJ,
DamleHuse}. For the golden chain, the mixed fixed point also has a
diagrammatic representation in terms of random trivalent graphs.
The mixed fixed point is the first non-singlet {\it stable}
infinite randomness fixed point to be discovered.

Another important character of this new IR fixed point is its
entanglement entropy. Both infinite randomness fixed points
exhibit the characteristic $\log L$ scaling:
 \be
S = \frac{1}{3} \, c_{eff}^{mixed} \, \ln L.
\ee
The coefficient in front of the $\ln$ for the pure AFM
chain was computed in Ref. \onlinecite{YB}, where it was found that it reflects the
quantum dimension of the Fibonacci anyons: $c_{eff}^{RS}=\ln \tau=0.481$.  The entropy scaling
calculation in the mixed phase is more intricate owing to the
complicated trivalent graph nature of the ground state.  We found the
effective central charge, $c_{eff}^{mixed}$, to be:
\be
c_{eff}^{mixed}=3\cdot 0.234=0.702.
\ee
Since this result was obtained through a combination of numerical and
analytical methods, it is hard to gain an intuitive understanding of
the numerical result. Nevertheless, it is interesting to compare it to
its pure-system analog, and to the effective central charge of the
random singlet phase. It is most likely that the mixed IR phase is
also the terminus of flow from the ferromagnetic pure fibonacci
chain. The central charge of the critical FM golden chains was
determined in Ref. \onlinecite{Qpeople1} to be
$c=4/5=0.8>c_{eff}^{mixed}$. Hence the effective central charged
dropped along the flow. Comparing our result, though, to the central
charge in the random singlet phase immediately reveals that the
effective central charge {\it increased} in the strong-randomness RG
flow from the random singlet phase to the mixed IR phase. Thus the
suggestion that strong-randomness flows may have a c-theorem
associated with them is contradicted.

This result is rather novel, and it is worth mentioning that one
can see that it is true without having to do the full calculation
of the entanglement entropy in the mixed FM/AFM phase.  To see
this, one first of all notices that the average $<l>$ between AFM
decimations, equal to $3/2$ (eqn. \ref{expl}), is half of that in
the AFM phase.  This means the treelike structures form twice as
fast as the singlets in the AFM phase.  Now, to compare
coefficients in front of $\ln L$ one must multiply by the energy
length scaling exponent, which is $1/3$ in the mixed phase and
$1/2$ in the AFM phase.  So the number of treelike structures
forming over the endpoints is $4/3$ times the number of singlets
in the forming in the AFM phase.  However, with probability $1/2$
the treelike structure is simply a singlet (AFM decimation
following another AFM decimation), and with probability $1/4$ it
is simply a tree on 3 sites (one FM decimation between the AFM
decimations), whose contribution to the entropy is the same as a
singlets.  Thus the treelike structures contribute at least $3/4$
as much entanglement entropy as singlets, and given that there are
$4/3$ times as many of them as singlets in the AFM phase, the
mixed phase has at least as much entropy as the AFM phase.  The
fact that the more complicated trees have a nonzero contribution
immediately shows that the entropy of the mixed phase is in fact
higher than that of the AFM phase.

\begin{figure} \vspace{5mm}
\includegraphics[width=8cm]{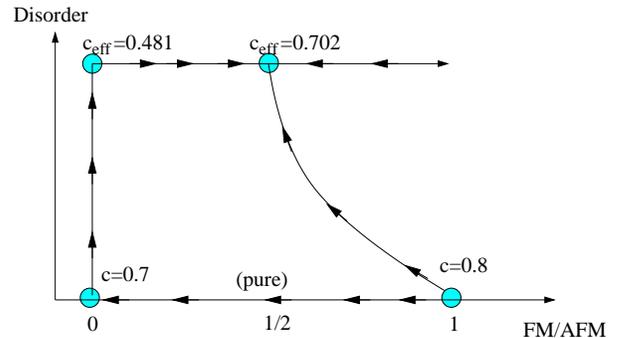}
\caption{Flow diagram of the pure and disordered golden chain. In the pure chain, assuming no intervening fixed points exist, the
  FM fixed point is unstable to flow to the AFM fixed point, as
  inferred from the Zamolodchikov c-theorem. In the disordered chain,
  however, the flow is in the opposite direction, with the mixed
  FM/AFM phase, which is most likely the terminus of the flow from the
  pure FM phase, being stable relative to the random singlet phase,
  which is the result of disordering the pure AFM phase. The fixed
  point (effective) central charges are also
  quoted. \label{fullflow}}
\end{figure}

Now, when we contrast the central charges of the of the two
critical phases of the pure chain, $c^{AFM}=0.7$, and $c^{FM}=0.8$, we
find that by the Zamolodchikov c-theorem,\cite{Zamolodchikov} the AFM phase must be a
stable phase with respect to the FM one, unless another critcal point
appears in between, which we speculate is unlikely. On the
other hand, the flow in the random golden chain is the opposite: the
mixed FM/AFM phase is stable for essentially all chain coupling
distributions, except for the point in which all couplings are
antiferromagnetic. This situation is summarized in
Fig. \ref{fullflow}.

Most importantly, we also observed in this paper the close connection
between a fusion algebra and real-space RG. This connection implies
that essentially all strongly disordered phases of non-abelian chains
will be of the infinite randomness class. In the future we intend to
analyze the random phases of non-abelian chains with different fusion
algebras. While this research is intended to expand our understanding
of random non-abelian systems, it may also lead to the discovery of
new infinite-randomness phases and universality classes, beyond the
permutation-symmetric sequence of Damle and Huse \cite{DamleHuse}.

\acknowledgments

We are indebted to A. Kitaev, J. Preskill,  S. Trebst, and P. Bonderson
for illuminating discussions. We would like to especially thank K. Yang
for his contributions to this project. G.R. and L.F. acknowledge
support from NSF Grant No. PHY-0456720.  N.E.B. acknowledges support from US DOE Grant
No. DE-FG02-97ER45639.  J.M. acknowledges support from DMR-0238760.  We would also like to acknowledge the KITP, UCSB, for their hospitality.

\bibliography{fusionrefs}
\end{document}